# Endoscopic Fourier-transform infrared spectroscopy through a fiber microprobe

Jaehyeon Kim,[1,2,*] Yue Tian,[1,2,*] Guanhua Qiao,[1,2] Julinna Abulencia Villarta,[1,2] Fujia Zhao,[1,2] Andrew He,[1,2] Ruo-Jing Ho,[3,4] Haoran Liu,[1,2] Rohit Bhargava,[3,4,5,6,7,8,9] and Yingjie Zhang[1,2,3,a)]

## AFFILIATIONS

[1]Department of Materials Science and Engineering, University of Illinois, Urbana, Illinois 61801, USA

[2]Materials Research Laboratory, University of Illinois, Urbana, Illinois 61801, USA

[3]Beckman Institute for Advanced Science and Technology, University of Illinois, Urbana, Illinois 61801, USA

[4]Department of Bioengineering, University of Illinois, Urbana, Illinois 61801, USA

[5]Department of Electrical and Computer Engineering, University of Illinois, Urbana, Illinois 61801, USA

[6]Cancer Center at Illinois, University of Illinois, Urbana, Illinois 61801, USA

[7]Department of Mechanical Science and Engineering, University of Illinois, Urbana, Illinois 61801, USA

[8]Department of Chemical and Biomolecular Engineering, University of Illinois, Urbana, Illinois 61801, USA

[9]Department of Chemistry, University of Illinois, Urbana, Illinois 61801, USA

[*]These authors contributed equally to this work.

a) Author to whom correspondence should be addressed: yjz@illinois.edu

## ABSTRACT

Fourier-transform infrared spectroscopy (FTIR) is a powerful analytical method for not only the chemical identification of solid, liquid, and gas species, but also the quantification of their concentration. However, the chemical quantification capability of FTIR is significantly hindered when the analyte is surrounded by a strong IR absorbing medium, such as liquid solutions. To overcome this limit, here we develop an IR fiber microprobe that can be inserted into liquid medium, and obtain full FTIR spectra at points of interest. To benchmark this endoscopic FTIR method, we insert the microprobe into bulk water covering a ZnSe substrate and measure the IR transmittance of water as a function of the probe–substrate distance. The obtained vibrational modes, overall transmittance vs z profiles, quantitative absorption coefficients, and micro z-section IR transmittance spectra are all consistent with the standard IR absorption properties of water. The results pave the way for endoscopic chemical profiling inside bulk liquid solutions, promising for applications in many biological, chemical, and electrochemical systems.

## I. INTRODUCTION

The mid-IR spectral range, approximately 4000–400 $cm^{-1}$ (2.5–25 μm), features rich vibrational states of many chemical substances and functional groups.[1–5] Therefore, mid-IR spectroscopy has been widely used for chemical fingerprinting and quantifications. The invention of FTIR has further enabled higher signal-to-noise ratio (SNR) and faster acquisition rate compared to dispersive spectrometers, thus finding applications as standard chemical analysis tools in research labs, industries, and government agencies worldwide.[6–8] However, unlike visible and near-IR light that easily penetrates through a large range of liquid, gas, and transparent solid media, many liquid species strongly absorb mid-IR light, making it challenging to perform mid-IR spectroscopy inside bulk liquids.[9–14] For example, visible light can easily penetrate through ~10 meters of water, while mid-IR beams are strongly attenuated within the scale of 10 microns.[15] As a result, the chemical identification and quantification of substances in aqueous solutions, such as live biological cells, is highly challenging.

A few approaches have been taken to overcome the strong background absorption of bulk liquid and obtain mid-IR spectrum inside liquid environments. One of the most widely used methods is attenuated total reflectance (ATR)-FTIR, which measures the near-surface liquid absorption through the ATR process.[5,11,12,16–18] However, the penetration depth of IR light is in the scale of only ~1 μm from the ATR crystal surface,[5,12,19,20] preventing the detection of species farther away inside the liquid. In addition, the excess light attenuation in the ATR accessory and the wavelength-dependence of penetration depth brings extra challenges in the precise concentration quantification of the near-surface substances. A similar yet more advanced approach is to couple IR light from the back-side of an IR-transparent solid to an atomic force microscopy (AFM) probe, thus achieving nanoscale in-plane resolution of the solid-liquid interface.[21–27] Nevertheless, the measured signal is further confined to within nanometers from the solid surface, and the quantification of IR absorption becomes even more challenging due to the convoluted coupling with the probe. Another approach is to use a liquid cell where a thin layer of analyte-containing liquid is sandwiched between two IR windows.[11,12,28–30] While the liquid thickness can be reduced down to sub-10 μm using this design, each liquid cell is only applicable to a limited range of samples (e.g., biological cells with specific size). Also, the confinement of liquid within the two IR windows strongly modifies the local microenvironment, and may perturb the inherent biological, chemical, and electrochemical processes.[28,31] A fourth strategy is to use emerging tunable lasers such as quantum cascade lasers and free-electron lasers. Harnessing the strong intensity of these light sources, a mid-IR path length larger than 100 μm has been achieved.[32] However, to date these tunable lasers can only produce a narrow mid-IR spectral range (a small fraction of the full mid-IR band offered by the FTIR source). In addition, the high laser power can lead to local heating effects which may perturb or damage sensitive samples.[33]

Here we take a new approach to overcome the background absorption problem of mid-IR spectroscopy—IR fiber probes. Initially developed a few decades ago, the first batches of mid-IR fibers had a significant loss larger than ~10 dB/m, which prevents their applications in communications and IR spectroscopy.[34] After significant efforts on materials processing, now the mid-IR fibers have been tremendously improved, reaching a loss less than 0.5 dB/m.[35–37] Such low loss ensures high-throughput transmission of mid-IR light through the fiber. Therefore, these fibers are ready to be used in liquid environments for targeted delivery of IR light to the desired spots for local spectroscopy measurements. Among the existing mid-IR fiber technologies, $As_2Se_3$ is particularly suitable for IR spectroscopy applications, due to their low optical loss,[35,36,38] nearly constant transmittance in the 4000–1000 cm$^{-1}$ spectral range,[35,39,40] and high chemical stability.[41,42]

Fiber-based microscopy has been previously demonstrated in the form of near-field scanning optical microscopy (NSOM).[43,44] Using tapered fiber probes to deliver and/or collect light, NSOM has achieved deep sub-wavelength resolution. However, its application is still limited to the visible and near-IR spectral range and to imaging in air.

Building upon the existing progress on $As_2Se_3$ fibers, here we report the fabrication of fiber microprobes to controllably access targeted regions inside liquid environments, and perform localized FTIR spectroscopy (Fig. 1). We envision this endoscopic IR approach and its future improvements to be broadly applicable to many systems, including biological live cells,[45–47] electrochemical systems (batteries, fuel cells, electrolyzers, etc.),[48–53] and heterogeneous catalysis processes (hydrogenation, $H_2O_2$ synthesis, etc.).[54–57] While there are many promising applications enabled by our developed IR microprobe, it is crucial to first benchmark the precise position-controlled spectroscopy capabilities. For this purpose, we measure the transmission spectra of pure water near an IR substrate, ZnSe, with varying probe–substrate distance ($z$). The results are fully consistent with the well-known Beer-Lambert law, the quantitative absorption coefficient ($\alpha$) of water, and the uniform distribution of water near the substrate with micron-scale $z$ resolution. With further developments, we expect to achieve full 3D spectromicroscopy capabilities in liquid environments.

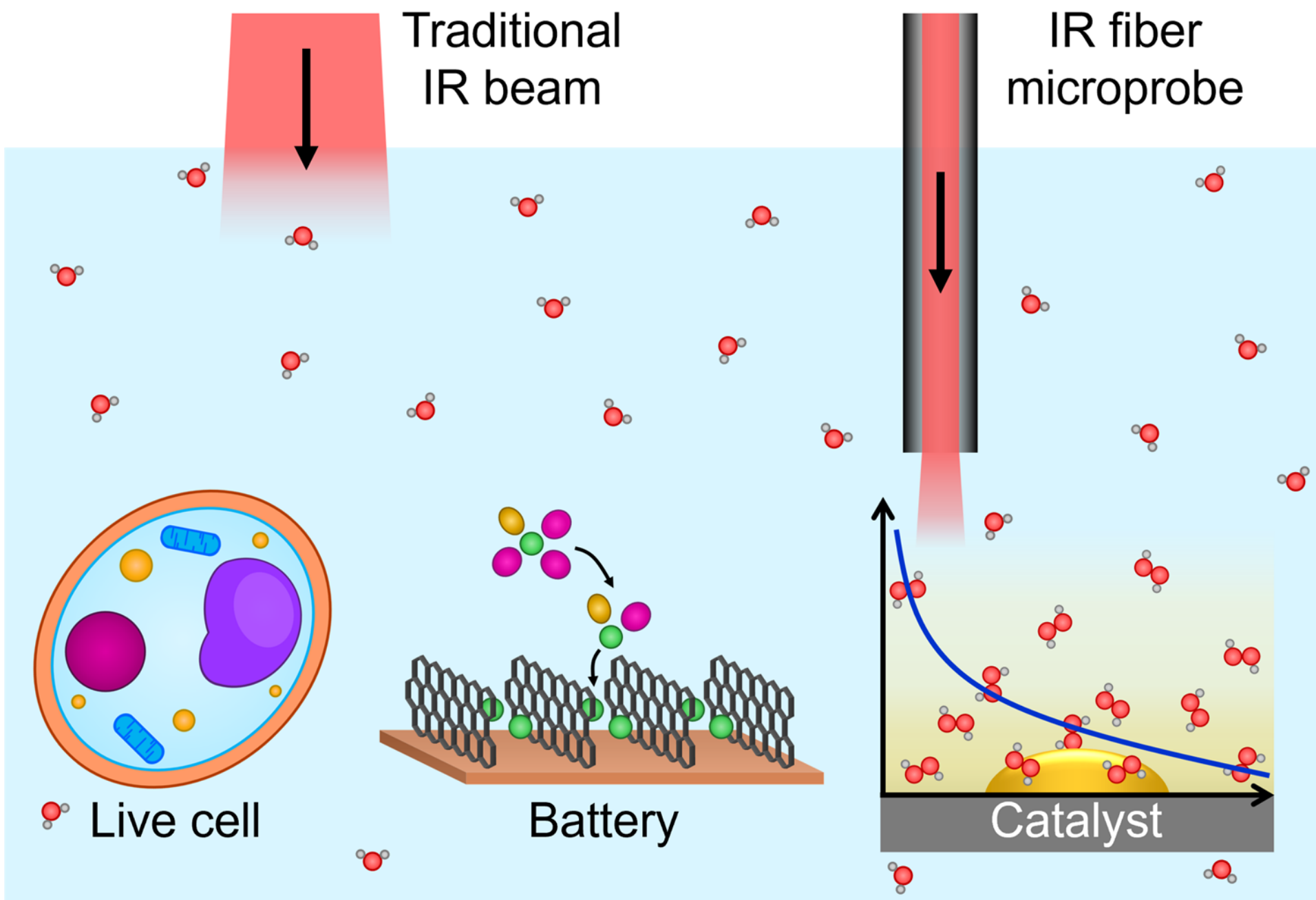

**FIG. 1.** Motivation and scope of applications of the FTIR microprobe method.

## II. INSTRUMENT DEVELOPMENT

### A. Fiber microprobe fabrication

We acquired $As_2Se_3$ optical fiber (IRF-Se-100R) from IRflex Corporation. These fibers have a spectral transmission range of 1.5–10 μm (6667–1000 $cm^{-1}$), with a specified loss near 0.5 dB/m. The core and cladding outer diameter of the fiber is 100 μm and 170 μm, respectively. The core/cladding is coated with polyacrylate (PA) with 90 μm thickness, resulting in a total diameter of 350 μm. The core refractive index is 2.7, and the numerical aperture (NA) is 0.27.

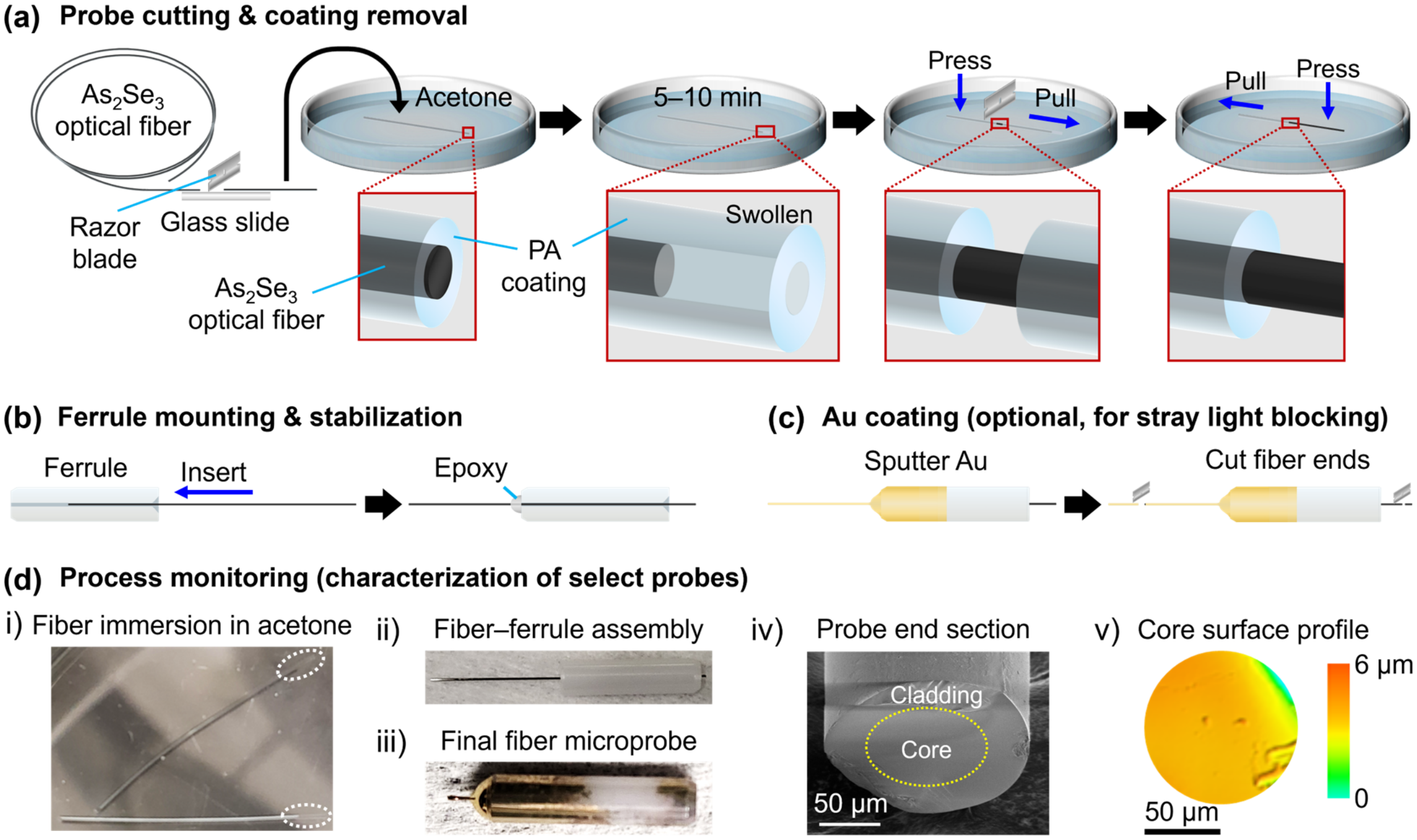

**FIG. 2.** Fiber microprobe fabrication process. Schematics of (a) the fiber cutting and PA coating stripping procedures, (b) fiber–ferrule assembly process, and (c) Au coating for some probes to reduce stray light. (d) Examples of probe characterization as part of the process monitoring, including: (i) photo of PA-coated $As_2Se_3$ fiber pieces immersed in acetone, with white dotted ovals marking the swollen PA protruding out of the $As_2Se_3$ core/cladding; (ii) photo of a fiber–ferrule assembly; (iii) photo of a final read-to-use fiber microprobe (the fiber protrusion at the right end is too short to be visible); (iv) SEM of a probe end section, with core and cladding marked; and (v) optical profilometry of a core area of the probe end surface (Sa~0.3 μm).

The overall fiber microprobe fabrication process is shown in Fig. 2. We begin by cutting a segment (typically 2–2.5 cm long) of the purchased $As_2Se_3$ fiber using a razor blade. The PA coating, important for preserving the internal structure of long fibers for optical communication applications, is soft and not ideal for the purpose of precise position control down to at least ~1 μm. Therefore, we proceeded to remove the PA coating of the fiber [Fig. 2(a)]. To this end, we first immersed the short fiber piece in a petri dish filled with acetone (≥ 99.5 %, Macron Fine Chemicals, MK244016, or ≥ 99.5%, Fisher Chemical, A18P-4) for 5–10 minutes to soften the PA coating. After this process, typically the fiber will swell and one end will protrude out of the $As_2Se_3$ core/cladding [Fig. 2(d)(i)]. Then, we used the blade to make a cut to the fiber coating (while still immersed in acetone). The cut should be gentle enough to avoid scratching the core and cladding. The PA coating layer (still in acetone) was then stripped off the core/cladding from the two sides consecutively by hand (while wearing gloves). We then took the bare fiber piece out of acetone and examined the straightness. If a bending angle larger than ~3° was observed, we will perform a straightening process as illustrated in Fig. S1. A straight fiber will enable a higher transmittance using the FTIR setup described later.

Since the bare fiber is slim and fragile, it needs to be stabilized first, before it can be used as a position-controlled endoscopy probe. For this purpose, we inserted the bare fiber piece into the bore of a ceramic ferrule (Thorlabs, Inc., CF230-10, inner diameter: 231 μm), and sealed one end using epoxy (Devcon 5 Minute Epoxy) [Fig. 2(b)]. The fiber was positioned so that its IR input end protruded from the concave side of the ferrule with a length of 1–2 mm, and the output end extended out of the flat side of the ferrule [Fig. 2(d)(ii)]. The epoxy serves the purposes of mechanically fixing and stabilizing the fiber probe, as well as reducing possible stray lights transmitting in the gap between the fiber and inner wall of the bore.

As an effort to further minimize the stray light, for some of the fiber probes, we coated a layer of Au on the fiber–ferrule assembly [Fig. 2(c)]. This was achieved using AJA Orion-8 Magnetron Sputtering System, with the fiber fixed vertically pointing towards the Au sputtering source, depositing 100 nm thick gold film under 65 W in 3 mTorr Ar atmosphere. For some probes, a 10 nm Cr layer (power: 100 W, pressure: 3 mTorr) was deposited right before Au coating. The Au coating serves as an anti-reflection layer that may further block adventitious light emission from undesired locations such as the cladding side wall and the gap between the cladding and ferrule bore (regions not covered by sufficiently thick epoxy). After sputtering, the two ends of the fiber were cut again to expose clean, Au-free cross-sections.

For endoscopy applications, the quality of the end surfaces of the fiber probe is crucial. They need to be sufficiently flat to both enable efficient coupling of IR light in and out of the probe, and to ensure precise control of the volume of probed regions in liquid. After fabricating each fiber–ferrule assembly, we performed optical profilometry measurements (using Keyence VK-X1000 3D Laser Scanning Confocal Microscope) to examine the surface roughness, and extract the arithmetic mean height (Sa) of the core area. Unless the desired roughness was achieved, we repeated the blade cutting of the fiber ends to eventually reach an Sa close to or lower than 1 μm. Examples of a photo of a finished fiber probe, scanning electron microscopy (SEM, FEI Helios NanoLab 600i) of a fiber end section, and optical height profile of a core region are shown in Fig. 2(d)(iii–v), respectively.

### B. Construction of the FTIR endoscopy setup

After fabricating the ferrule-mounted fiber microprobe, we proceeded to build the setup for mid-IR endoscopy measurements. To facilitate the broad dissemination of this method in the future, we constructed a modular setup. The core opto-mechanical components were assembled onto a small optical breadboard (Thorlabs, MB8, 8" × 8" × 1/2") [Figs. 3(a) and 3(b)], which can then be readily mounted into many existing mid-IR spectrometers to enable endoscopic measurements.

The setup is designed to enable two key functions: light coupling through the fiber probe, and micro-position control of the sample [Fig. 3(b)]. To couple free-space IR light source into the fiber probe, we first used a broadband ZnSe focusing objective lens (Edmund Optics, 88-448, 0.08 NA, 18 mm focal length (FL)) to focus the IR light to a small spot comparable to the fiber diameter. The objective lens was mounted on a Z-axis translation mount (Thorlabs, SM1ZA) to allow position adjustments. Since the fiber core diameter is only 100 μm, light coupling into the fiber requires micro-position control of the fiber–ferrule assembly. This was achieved by securing the fiber–ferrule assembly into a ferrule adapter plate (Thorlabs, SM1FCM, made of Al 6061-T6) that serves as a probe holder (similar to the roles of probe holders in most existing scanning probe microscopy setups[58–62]), which was further fixed on an XY translator (Thorlabs, ST1XY-D) [Fig. 3(c)]. This microprobe stage enables micron-scale position adjustments of the fiber probe along the $x$ and $y$ directions. After the IR light transmits through the fiber probe and the sample, it is refocused by a ZnSe plano-convex lens (Thorlabs, LA7656-E4, 50.1 mm FL) before reaching the IR detector. This plano-convex lens was mounted into a lens tube (Thorlabs, SM1L03), which was further attached to a kinematic mirror mount (Thorlabs, KM100T) to enable angle adjustments.

The simple, transmission-based optical design facilitates the measurement of a large variety of samples. Here, to demonstrate the instrument capabilities, we use a ZnSe window (Thorlabs, WG71050, diameter: 25.4 mm, thickness: 5.0 mm, roughness: <1 μm) coated with an aqueous droplet as the sample [Figs. 3(b), 3(d), 3(e)]. The ZnSe window was secured into a mirror mount (Thorlabs, FMP1) and then fixed to an XYZ translation stage (Thorlabs, MT3A) to achieve 3D position control with ~1 μm precision. Through manual adjustments, we were able to move the substrate to align and position the targeted areas of the sample to the fiber microprobe and achieve mid-IR endoscopy measurements [Figs. 3(d) and 3(e)]. FTIR measurement of the ZnSe window in air revealed nearly constant transmittance close to 80% in the wavenumber range of 1000–4000 $cm^{-1}$ (Fig. S2). Although the ZnSe substrate was positioned vertically in our design, an aqueous droplet placed on the ZnSe surface typically remains stable without falling down, due to surface tension [Fig. 3(e)]. Without the addition of more water/aqueous solution, one droplet (20–30 μL) usually evaporates after at least ~1 hour, which is sufficient for collecting a complete series of endoscopic FTIR spectra at different locations.

With the small size and adjustable positions of individual parts, our constructed opto-mechanical module can be readily integrated into most commercial and home-built mid-IR spectrometers. As a first demonstration, we mounted this module into a widely used commercial FTIR spectrometer, Thermo Nicolet iS50, which has a Polaris mid-IR source with a spectral range of 10–9600 $cm^{-1}$, and a liquid nitrogen-cooled MCT-A detector sensitive to the spectral range of 600–11700 $cm^{-1}$ [Figs. 3(a) and 3(b)]. The IR beam was illuminated from the right port of the sample compartment, and finally collected by the left port to the detector. The opto-mechanical module components were aligned to maximize the IR throughput, before each FTIR endoscopy measurement. The alignment included centering of all optical components, making lateral focus adjustments for the objective lens and the plano-convex lens, and angular adjustments for the objective lens, the ZnSe window, and the plano-convex lens. This alignment process was performed using an empty ferrule (with no fiber inserted). Once completed, the fiber–ferrule assembly was used, and optimal alignment was determined by maximizing the peak intensity of the FTIR interferogram.

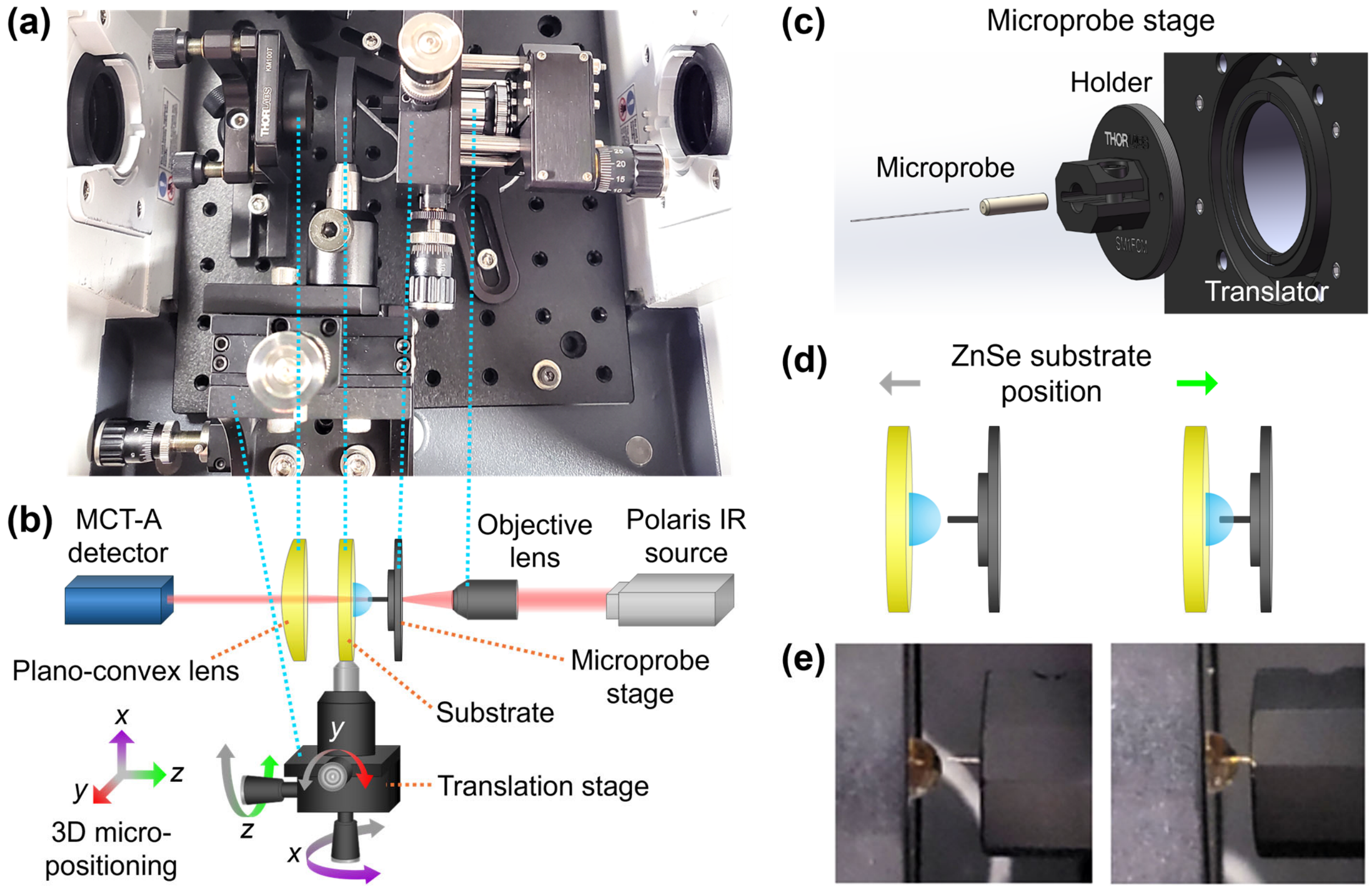

**FIG. 3.** Construction of the mid-IR endoscopy setup. (a) Photo and (b) schematic of the microprobe-based FTIR endoscopy setup. (c) Schematic of the microprobe stage (reproduced with permission from Thorlabs, Inc.). (d) Schematic and (e) photos of an IR fiber microprobe performing endoscopy of a liquid droplet (water in (e)).

## III. INSTRUMENT BENCHMARKING

To benchmark the capabilities of the FTIR endoscopy setup, it is desirable to choose a sample with a well-known quantitative mid-IR absorption spectrum, and widely used in many research and application areas. To this end, pure water is an ideal choice, as it is the source of life and medium for many chemical and biological processes.[10,30,63–66] In all the measurements, we recorded the raw single beam spectrum (direct Fourier transform of the interferogram) of each sample/configuration, before further processing to obtain the transmittance.

We began the experiment by collecting a single beam spectrum in air ($I_{air}$), using the setup shown in Figs. 3(a) and 3(b) where the sample was a bare ZnSe window with no liquid droplet [Fig. S3(a)]. As shown in Fig. 4(a), the air spectrum has an overall shape characteristic of blackbody radiation, as expected from the mid-IR source. On top of the overall background, we observed a series of sharp peaks due to the environmental species, including: peaks between ~1400–1900 $cm^{-1}$ due to H–O–H bending of water vapor,[67,68] a doublet peak at ~2350 $cm^{-1}$ from asymmetric stretching of atmospheric $CO_2$,[67–70] ~2800–3000 $cm^{-1}$ from C–H stretching of airborne hydrocarbons (either in air or at surfaces of optics parts),[71] and ~3500–3950 $cm^{-1}$ due to O–H stretching of water vapor.[67,68]

We then drop cast 20–30 μL of Milli-Q water (18.2 MΩ·cm at 25 °C; Synergy UV water purification system, MilliporeSigma) on the front surface of the ZnSe substrate, and moved the microprobe into the water to collect FTIR spectrum. For short probes extending ~1–2 mm out of the probe holder, the water droplet usually forms a meniscus between the ZnSe substrate and the holder's front surface (Fig. S3), which further stabilizes the liquid configuration.

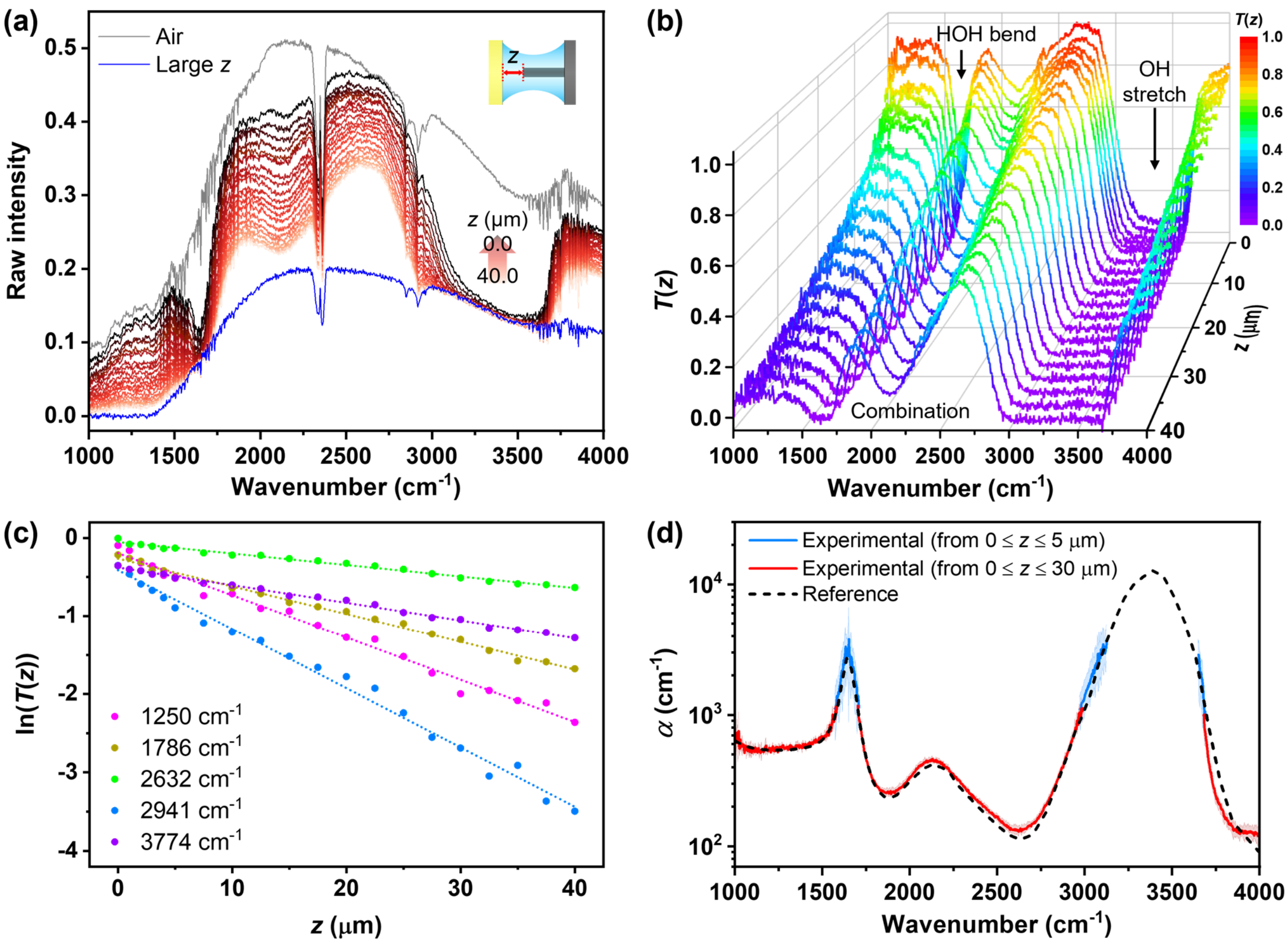

**FIG. 4.** Benchmarking the endoscopic FTIR technique by measuring in water. (a) Single beam spectra of air and water at different $z$. $z$ was decreased from 40.0 μm to 5.0 μm using a step size of 2.5 μm, and from 5.0 μm to 0 with 1.0 μm step. The inset schematic illustrates the experimental configuration including the probe, ZnSe substrate, and water. (b) IR transmittance spectra of water at different $z$, obtained by applying Eq. (1) to the data in (a). IR resolution = 4 cm$^{-1}$, data spacing = 1.928 cm$^{-1}$, the number of scans = 8, gain = 8, optical velocity of the moving mirror = 0.1581 cm/s, beam aperture = 230, KBr beamsplitter was used. (c) Natural log of the IR transmittance of water at different $z$, together with their linear fitting, for select wavenumbers. (d) Absorption coefficient spectrum of water extracted from the linear fits of $ln(T(z))$ vs $z$ at each wavenumber using Eq. (2), where the $z$ range for the fit is 0–30 μm (red solid line) or 0–5 μm (blue solid line). The darker core and lighter shaded regions represent the average and error bar (standard deviation), respectively, obtained from the three different data sets (Fig. S12). Reference spectrum (black dashed line) is adapted with permission from Appl. Opt. 12, 555 (1973).[15] Copyright Optical Society of America.

A powerful capability of our microprobe-based FTIR measurement is the precise control of the probe position relative to the sample, which enables quantitative transmission measurements through precisely modulated volumes of liquid. For water, the $\alpha$ in the mid-IR range is within ~ $10^2$–$10^4$ cm$^{-1}$,[15] corresponding to a penetration depth ($1/\alpha$) of ~ 1–100 μm. Therefore, at a $z \gg$ 100 μm, with the probe immersed in water, the transmittance through the probe–substrate gap should be zero, although the single beam intensity may be non-zero if stray lights are present. Due to these considerations, we first brought the ZnSe window to more than ~0.3 mm from the probe (still inside the water droplet) to collect the single beam spectrum ($I_{large\ z}$), which we regard as the stray light background [Fig. 4(a)].

We then moved the probe closer to the substrate and performed $z$-profiling, i.e., a series of spectra at different $z$ (between 0–40 μm). As shown in Fig. 4(a), at such small $z$, in most of the mid-IR spectral range, the single beam intensity ($I(z)$) is higher than that of the stray light background obtained at $z$>0.3 mm ($I_{large\ z}$), due to the extra IR transmission through the water filling the probe–substrate gap. At smaller $z$, $I(z)$ becomes higher, as a result of stronger IR transmission through a thinner region of water in the probe–substrate gap. Eventually, at $z$~0 μm, $I(z)$ becomes close to that measured in air, except near the H–O–H bending mode at ~1650 cm$^{-1}$ and O–H stretching mode at 2900–3700 cm$^{-1}$. These raw detector signals already reveal the $z$-modulation of the water absorption effects. Note that $z$~0 μm is assigned as the $z$ position below which negligible changes in the single beam intensity can be observed (Fig. S4). The assigned zero point is likely the point at which the probe surface first became in contact with the ZnSe substrate. Although these two surfaces were in contact, due to the micron-scale height variations of the probe surface [Fig. 2(d)(v)], there can be many micro-puddles of water trapped between the probe and substrate (Fig. S5). These residual amounts of water likely resulted in a lower IR transmittance compared to that in air in the spectral range for water vibration [Fig. 4(a)]. After the probe retracted from the ZnSe substrate, no damage to the probe structure was observed [Fig. S3(d)].

To quantify the transmittance of water within the probe–substrate gap, we use the following formula:

$$T(z) = \frac{I(z) - I_{large\ z}}{I_{air} - I_{large\ z}}. \quad (1)$$

The rationale is that, the stray light background $I_{large\ z}$ is likely the same for liquid and air measurements, and are in parallel with the "true signal" transmitted between the probe and the substrate (Fig. S6). Therefore, $I(z) - I_{large\ z}$ and $I_{air} - I_{large\ z}$ are approximately the transmitted and incident light intensity through the water in the probe–substrate gap, respectively, and their ratio is the transmittance. Experimentally, we verified that $I_{air}$ and $I_{large\ z}$ are $z$-independent within up to a few mm distance (Figs. S7 and S8), further validating Eq. (1).

Using Eq. (1), we extracted the IR transmittance vs wavenumber at a series of $z$ through water, as shown in Fig. 4(b). These results are consistent with the standard transmittance spectra of bulk water, with characteristic features including the H–O–H bending mode at ~1650 cm$^{-1}$, combination mode (a combination of H–O–H bending and libration) at ~2130 cm$^{-1}$, and O–H stretching mode at ~2900–3700 $cm^{-1}$.[72–74] Remarkably, the sharp peaks from gaseous and airborne species in the single beam spectrum [Fig. 4(a)] were mostly gone in the water transmittance spectrum, further proving the effectiveness of Eq. (1). As an example, the zoomed-in view of the hydrocarbon C–H stretching spectral range is shown in Fig. S9. From Fig. 4(b), we can see that the overall transmittance increases at smaller $z$, except near the center of the O–H stretching peak where the transmittance is below the detection limit due to the strong absorption. With further improvements in the microprobe fabrication methods, we may be able to achieve a flatter probe surface, which can reduce the amount of residual water at zero probe–substrate distance and thus enable the detection of the water transmittance throughout the whole spectral range.

Since water is a homogeneous medium, its IR transmission is expected to follow the well-known Beer-Lambert law, which, in the natural log form, is:

$$\ln\big(T(z)\big) = -\alpha z, \tag{2}$$

where $\alpha$ is the absorption coefficient that is wavenumber-dependent. From ray tracing simulation, we observe that the divergence of the IR beam diameter is within 10% in the $z$ span of 0–40 μm (Fig. S10), indicating that Eq. (2) should be applicable within this range. To determine whether our $z$-profiling results indeed agree with the Beer-Lambert law, we plotted $\ln\big(T(z)\big)$ vs $z$ at a series of wavenumbers, and performed linear fits. As shown in Fig. 4(c), we observed the linear dependence, revealing that $T(z)$ does decrease exponentially vs $z$. From the linear fits, we extracted $\alpha$ from the slope, and plotted it vs wavenumber [Fig. 4(d)]. Considering the differences in penetration depth at different wavenumbers, we used different $z$ ranges for linear fitting and $\alpha$ extraction: 0–30 μm for wavenumbers with an expected penetration depth > 10 μm, and 0–5 μm for wavenumbers with an estimated penetration depth within 10 μm. Near the apex of the H–O–H stretching mode (3125–3650 $cm^{-1}$), the known $\alpha$ is above ~$3\times10^{3}$ $cm^{-1}$, corresponding to a penetration depth within ~3 μm. Since our measured transmittance is effectively zero within the noise limit in this highly absorptive wavenumber range, due to the micron-scale roughness of the probe, we were not able to extract an $\alpha$ value within 3125–3650 $cm^{-1}$. Except this range, for all other wavenumbers we observe nearly perfect overlap with the reference spectra [Fig. 4(d)].[15] This reveals the remarkable precise quantification capabilities of our developed endoscopic FTIR method.

The reproducibility of the water $z$-profiling experiment and the data processing methods were further verified in two other data sets shown in Figs. S11 and S12. Moreover, we examined the mechanical stability of our setup via optical imaging, and observed negligible position drift within the micron-scale optical resolution limit after 30 minutes, much longer than the ~40 s time required to collect a single spectrum (Fig. S13). These results thoroughly confirm the position-controlled endoscopic FTIR capabilities of our setup.

To further improve the setup, we investigated the origin of stray light and methods to eliminate it. Control single beam measurements with the IR source port blocked showed zero intensity (Fig. S14), revealing that the stray light we measured did not originate from any thermal background. Therefore, the stray light likely corresponds to the IR light emitted from the IR source but not transmitted through the end of the fiber probe (Fig. S6). To remove the excess signal, we used an IR-absorbing tape to cover nearly everywhere at the output end of the fiber probe holder except the fiber itself. The resulting water $z$-profiling measurements do confirm the stray light removal and the ability to precisely quantify the absorption coefficient spectrum of water (Fig. S15). As a next step, we will replace the tape method by filling the fiber–ferrule and ferrule–holder gaps with IR-absorbing material, to achieve a cleaner and more stable setup for liquid measurements.

Another area of improvement is the optical coupling efficiency of the fiber. As shown in Fig. S16, currently the measured single beam intensity through the $As_2Se_3$ fiber is only ~20% of that through a 100 μm pinhole. This low transmittance of the fiber is likely limited by the micron-scale roughness of the front and end surfaces. With further reduction of the fiber surface roughness (e.g. through polishing or focused ion beam cutting), we may increase the SNR, sensitive spectral range, as well as the detectable $z$ range of the FTIR endoscopy method.

## IV. TOWARDS QUANTITATIVE SPECTROMICROSCOPY IN LIQUID

The ability to measure FTIR transmission spectra at controlled positions opens the door for quantitative spectromicroscopy in liquid environments. To demonstrate this capability, we reckon that, in the above-described water $z$-profiling experiments, IR transmittance through a thin $z$ section of liquid can be quantitatively extracted. That is, in the section between $z_1$ and $z_2$ ($z_1 < z_2$), liquid transmittance is:

$$T(z_1, z_2) = \frac{T(z_2)}{T(z_1)}. \tag{3}$$

Note that, Eq. (3) is universally applicable, even if the liquid is inhomogeneous and does not follow Eq. (2). From the $T(z)$ results shown in Fig. 4(b), we further extracted the IR transmittance of thin sections of water using Eq. (3). The results, as shown in Fig. 5, reveal nearly identical transmittance spectra through the sections with the same thickness but different distance away from the ZnSe substrate.

More specifically, within the wavenumber range of 1800–2850 cm$^{-1}$, $\alpha$ is within ~ $10^2$–$5\times10^2$ cm$^{-1}$ [Fig. 4(d)], corresponding to a penetration depth of ~ 20–100 μm. In this spectral range, micro-sections with 5 μm thickness clearly resolved the water combination mode centered around ~2130 cm$^{-1}$ [Fig. 5(b)], while 1 μm sections are two thin to distinguish this mode from the high transmittance background [Fig. 5(c)]. In contrast, near the ~1650 cm$^{-1}$ H–O–H bending peak, $\alpha$ is above ~$10^3$ cm$^{-1}$ [Fig. 4(d)], corresponding to a penetration depth lower than ~10 μm. This strongly absorptive mode was clearly identified in the transmittance through 1 μm sections [Fig. 5(c)]. Also, for both the combination mode in Fig. 5(b) and the H–O–H bending mode in Fig. 5(c), while the overall section transmittance values $T(z_1, z_2)$ remain independent of the average distance from the ZnSe substrate (i.e., $(z_1 + z_2)/2$), the SNR decreases for the sections farther away from the substrate. This is a natural result of the lower overall transmittance ($T(z)$) at larger $z$ [Fig. 4(b)], leading to a lower SNR for both $T(z)$ and $T(z_1, z_2)$.

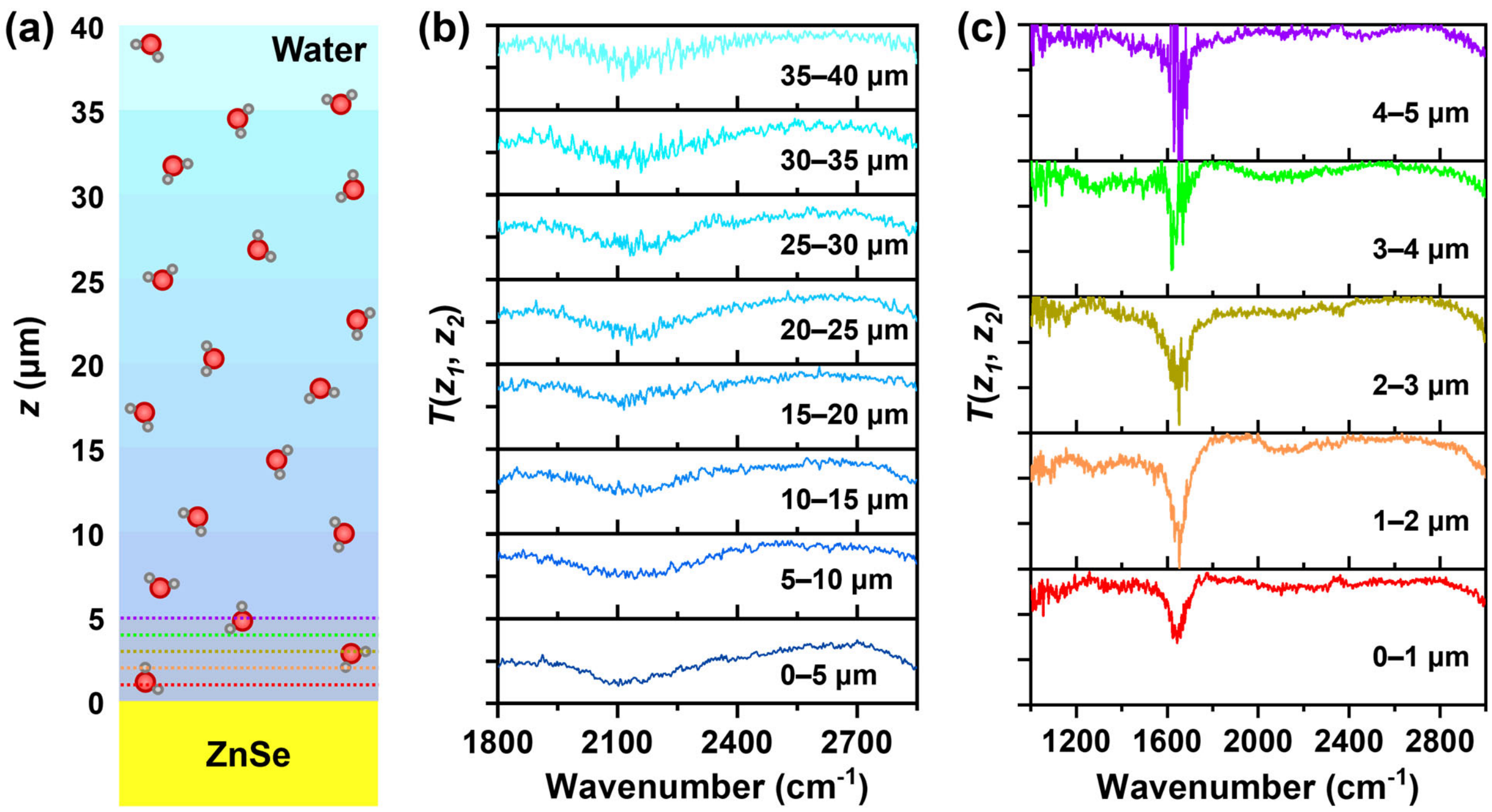

**FIG. 5.** Micro-section analysis of IR transmittance through water. (a) Schematic of different $z$ sections of water near the ZnSe substrate. The dotted lines differentiate the 1 μm sections in the $z$ range of 0–5 μm. (b) and (c) show the section transmittance $T(z_1, z_2)$ extracted from Fig. 4(b) using Eq. (3), with a section thickness of 5 μm and 1 μm, respectively. For each plot in (b), the y axis range is 0.6–1.0, while each panel in (c) has a y axis range of 0.4–1.0.

This $z$ sectioning capability will be highly valuable for two types of purposes. One is the detection and determination of concentration gradients of chemical species in the diffusion layers of liquid electrolyte during electrochemical and heterogeneous catalytic reactions, such as $CO_2$ reduction,[75,76] $H_2O_2$ synthesis,[77] and heterogeneous hydrogenation.[78] In these systems, the diffusion layer thickness near solid electrodes/catalysts are typically a few 10s to 100 μm,[75,77,79–81] which can be readily measured by our FTIR microprobe $z$-profiling method. Another application is the quantification and differentiation between the interfacial/interior species of a solid/biological cell and their surrounding liquid environment.[82,83] This is because the transmission spectra measured at $z$~0 corresponds to the species either at the very surface (within a thickness limited by the roughness of the microprobe and/or the substrate) or both at the surface and interior of the solid/cell system, depending on the measurement configuration of $I_{air}$ used in Eq. (1); in contrast, $T(z_1, z_2)$ at $z$>0 enables direct quantification of near-surface species.

Additionally, beyond the $z$ sectioning capabilities with micron-scale resolution, scanning along the $(x, y)$ directions will enable in-plane spectromicroscopy with a resolution down to the aperture opening (currently ~100 μm, limited by the core diameter of the fiber probe (Fig. S10). Both the $z$ and $(x, y)$ resolution may be further improved significantly by fabricating tapered probes, similar to the NSOM techniques that have been demonstrated for the visible and near-IR spectral range.[43,44]

## V. CONCLUSION

In conclusion, we have demonstrated a fiber microprobe-based method that has enabled endoscopic FTIR in liquid environments. This instrument currently has micro-positioning capabilities and has achieved ~1 μm $z$-resolution in chemical profiling, with enormous potential for further resolution improvement and for achieving full 3D spectromicroscopy. The method also has a larger spectral detection range compared to traditional liquid cell-based FTIR,[67] as well as higher sensitivity vs confocal Raman spectromicroscopy.[75] We envision a series of applications in biological, chemical, and electrochemical systems where liquid environments are essential yet difficult to access using other tools.

## SUPPLEMENTARY MATERIAL

The supplementary material includes photos of the fiber straightening process, IR transmittance spectrum of the ZnSe window, photos of the probe and sample configurations during endoscopic FTIR measurements, $z$ value calibration method, schematic illustrating that water can exist at $z$=0 μm due to the probe's surface roughness, schematic of the possible light paths in different experimental configurations, single beam spectra measured in air at different $z$, single beam spectra of water at different $z$ larger than 0.3 mm, $z$-profiling results in the spectral range for C–H stretching, estimated spatial profile of the IR beam transmitted out of the $As_2Se_3$ optical fiber, additional endoscopic FTIR measurements of water, absorption coefficient spectra of water extracted from the three individual endoscopic FTIR measurements, mechanical stability of the FTIR endoscopy setup, thermal background measurement of the FTIR endoscopy setup, origin and elimination of the stray light, and optical throughput measurement of the fiber microprobe.

## ACKNOWLEDGMENTS

J.K. and Y.Z. acknowledge the support by the Air Force Office of Scientific Research under Award No. FA9550-22-1-0014. Y.T., J.A.V., R.-J.H., R.B., and Y.Z. acknowledge the support from the National Science Foundation under Grant No. 2243257. G.Q., F.Z., A.H., H.L., and Y.Z. acknowledge the support from the Beckman Young Investigator Award provided by the Arnold and Mabel Beckman Foundation. The experiments were performed in part in the Materials Research Laboratory at the University of Illinois.

## AUTHOR DECLARATIONS

### Conflict of Interest

The authors have no conflicts to disclose.

### Author Contributions

**Jaehyeon Kim:** Data curation (supporting); Formal analysis (lead); Investigation (equal); Validation (equal); Visualization (lead); Writing - original draft (supporting); Writing - review & editing (equal). **Yue Tian:** Data curation (lead); Formal analysis (supporting); Methodology (lead); Investigation (equal); Resources (equal); Validation (equal); Writing - original draft (supporting); Writing - review & editing (supporting). **Guanhua Qiao:** Methodology (supporting); Investigation (supporting); Resources (equal); Writing - original draft (supporting); Writing - review & editing (supporting). **Julinna Abulencia Villarta:** Investigation (supporting); Validation (supporting); Visualization (supporting); Writing - original draft (supporting); Writing - review & editing (supporting). **Fujia Zhao:** Data curation (supporting); Methodology (supporting); Investigation (supporting); Resources (supporting). **Andrew He:** Methodology (supporting); Investigation (supporting); Resources (supporting). **Ruo-Jing Ho:** Methodology (supporting); Writing - review & editing (supporting). **Haoran Liu:** Methodology (supporting); Investigation (supporting); Resources (supporting). **Rohit Bhargava:** Funding acquisition (supporting); Writing - review & editing (supporting). **Yingjie Zhang:** Conceptualization (lead); Funding acquisition (lead); Project administration (lead); Supervision (lead); Validation (equal); Writing - original draft (lead), Writing - review & editing (equal). J.K. and Y.T. contributed equally to this work.

## DATA AVAILABILITY

The data that support the findings of this study are available from the corresponding author upon reasonable request.

## REFERENCES

[1] J. Haas and B. Mizaikoff, "Advances in mid-Infrared spectroscopy for chemical analysis," Annu. Rev. Anal. Chem. **9**(1), 45–68 (2016).
[2] M. Hlavatsch, A. Teuber, M. Eisele, and B. Mizaikoff, "Sensing liquid- and gas-phase hydrocarbons via mid-infrared broadband femtosecond laser source spectroscopy," ACS Meas. Sci. Au **3**(6), 452–458 (2023).
[3] V. A. Lorenz-Fonfria, "Infrared difference spectroscopy of proteins: From bands to bonds," Chem. Rev. **120**(7), 3466–3576 (2020).
[4] M. O. Guerrero-Pérez and G. S. Patience, "Experimental methods in chemical engineering: Fourier transform infrared spectroscopy—FTIR," Can. J. Chem. Eng. **98**(1), 25–33 (2020).
[5] I. A. Mudunkotuwa, A. A. Minshid, and V. H. Grassian, "ATR-FTIR spectroscopy as a tool to probe surface adsorption on nanoparticles at the liquid–solid interface in environmentally and biologically relevant media," Analyst **139**(5), 870–881 (2014).
[6] Z. Movasaghi, S. Rehman, and I. U. Rehman, "Fourier transform infrared (FTIR) spectroscopy of biological tissues," Appl. Spectrosc. Rev. **43**(2), 134–179 (2008).
[7] C. Corsi, "Infrared: A key technology for security systems," Adv. Opt. Technol. **2012**, 1–15 (2012).
[8] D. -W. Sun, *Infrared Spectroscopy for Food Quality Analysis and Control*, 1st ed (Academic Press, Burlington, MA, 2009).

[9] D. Brissinger, G. Parent, and P. Boulet, “Experimental study on radiation attenuation by a water film,” J. Quant. Spectrosc. Radiat. Transfer **145**, 160–168 (2014).
[10] K. Rahmelow and W. Hübner, “Infrared spectroscopy in aqueous solution: Difficulties and accuracy of water subtraction,” Appl. Spectrosc. **51**(2), 160–170 (1997).
[11] P. S. Fomina, M. A. Proskurnin, B. Mizaikoff, and D. S. Volkov, “Infrared spectroscopy in aqueous solutions: Capabilities and challenges,” Crit. Rev. Anal. Chem. **53**(8), 1748–1765 (2023).
[12] F. Zaera, “Probing liquid/solid interfaces at the molecular level,” Chem. Rev. **112**(5), 2920–2986 (2012).
[13] T. Buffeteau, J. Grondin, and J.-C. Lassègues, “Infrared spectroscopy of ionic liquids: Quantitative aspects and determination of optical constants,” Appl. Spectrosc. **64**(1), 112–119 (2010).
[14] J. M. Porter, J. B. Jeffries, and R. K. Hanson, “Mid-infrared absorption measurements of liquid hydrocarbon fuels near,” J. Quant. Spectrosc. Radiat. Transfer **110**(18), 2135–2147 (2009).
[15] G. M. Hale and M. R. Querry, “Optical constants of water in the 200-nm to 200-µm wavelength region,” Appl. Opt. **12**(3), 555 (1973).
[16] V. J. Ovalle, Y. -S. Hsu, N. Agrawal, M. J. Janik, and M. M. Waegele, “Correlating hydration free energy and specific adsorption of alkali metal cations during $CO_2$ electroreduction on Au,” Nat. Catal. **5**(7), 624–632 (2022).
[17] H. Zhang, B. Chen, T. Liu, G. W. Brudvig, D. Wang, and M. M. Waegele, “Infrared spectroscopic observation of oxo- and superoxo-intermediates in the water oxidation cycle of a molecular Ir catalyst,” J. Am. Chem. Soc. **146**(1), 878–883 (2024).
[18] G. Li, Z. -C. An, J. Yang, J. -H. Zheng, L. -F. Ji, J. -M. Zhang, J. -Y. Ye, B. -W. Zhang, Y. -X. Jiang, and S. -G. Sun, “Revealing surface fine structure on PtAu catalysts by an *in situ* ATR-SEIRAS CO-probe method,” J. Mater. Chem. A **11**(26), 14043–14051 (2023).
[19] H. Kaur, B. Rana, D. Tomar, S. Kaur, and K. C. Jena, “Fundamentals of ATR-FTIR spectroscopy and its role for probing in-situ molecular-level interactions,” in *Modern Techniques of Spectroscopy*, edited by D.K. Singh, M. Pradhan, and A. Materny, (Springer Singapore, Singapore, 2021), pp. 3–37.
[20] B. Byrne, J. W. Beattie, C. L. Song, and S. G. Kazarian, “ATR-FTIR spectroscopy and spectroscopic imaging of proteins,” in *Vibrational Spectroscopy Protein Research* (Elsevier, 2020), pp. 1–22.
[21] B. T. O’Callahan, K. -D. Park, I. V. Novikova, T. Jian, C. -L. Chen, E. A. Muller, P. Z. El-Khoury, M. B. Raschke, and A. S. Lea, “In liquid infrared scattering scanning near-field optical microscopy for chemical and biological nanoimaging,” Nano Lett. **20**(6), 4497–4504 (2020).
[22] H. Wang, J. M. González-Fialkowski, W. Li, Q. Xie, Y. Yu, and X. G. Xu, “Liquid-phase peak force infrared microscopy for chemical nanoimaging and spectroscopy,” Anal. Chem. **93**(7), 3567–3575 (2021).
[23] H. Wang, E. Janzen, L. Wang, J. H. Edgar, and X. G. Xu, “Probing mid-infrared phonon polaritons in the aqueous phase,” Nano Lett. **20**(5), 3986–3991 (2020).
[24] Y. -H. Lu, J. M. Larson, A. Baskin, X. Zhao, P. D. Ashby, D. Prendergast, H. A. Bechtel, R. Kostecki, and M. Salmeron, “Infrared nanospectroscopy at the graphene–electrolyte interface,” Nano Lett. **19**(8), 5388–5393 (2019).
[25] L. Xiao and Z. D. Schultz, “Spectroscopic imaging at the nanoscale: Technologies and recent applications,” Anal. Chem. **90**(1), 440–458 (2018).
[26] Y. Shan, X. Zhao, M. Fonseca Guzman, A. Jana, S. Chen, S. Yu, K. C. Ng, I. Roh, H. Chen, V. Altoe, S. N. Gilbert Corder, H. A. Bechtel, J. Qian, M. B. Salmeron, and P. Yang, “Nanometre-

resolved observation of electrochemical microenvironment formation at the nanoparticle–ligand interface," Nat. Catal. **7**(4), 422–431 (2024).
[27] E. Pfitzner and J. Heberle, "Infrared Scattering-Type Scanning Near-Field Optical Microscopy of Biomembranes in Water," J. Phys. Chem. Lett. **11**(19), 8183–8188 (2020).
[28] J. Doherty, Z. Zhang, K. Wehbe, G. Cinque, P. Gardner, and J. Denbigh, "Increased optical pathlength through aqueous media for the infrared microanalysis of live cells," Anal. Bioanal. Chem. **410**(23), 5779–5789 (2018).
[29] J. Doherty, A. Raoof, A. Hussain, M. Wolna, G. Cinque, M. Brown, P. Gardner, and J. Denbigh, "Live single cell analysis using synchrotron FTIR microspectroscopy: Development of a simple dynamic flow system for prolonged sample viability," Analyst **144**(3), 997–1007 (2019).
[30] H. Yang, S. Yang, J. Kong, A. Dong, and S. Yu, "Obtaining information about protein secondary structures in aqueous solution using Fourier transform IR spectroscopy," Nat. Protoc. **10**(3), 382–396 (2015).
[31] E. W. K. Young and D. J. Beebe, "Fundamentals of microfluidic cell culture in controlled microenvironments," Chem. Soc. Rev. **39**(3), 1036 (2010).
[32] T. F. Beskers, M. Brandstetter, J. Kuligowski, G. Quintás, M. Wilhelm, and B. Lendl, "High performance liquid chromatography with mid-infrared detection based on a broadly tunable quantum cascade laser," Analyst **139**(9), 2057 (2014).
[33] A. Schwaighofer, M. Brandstetter, and B. Lendl, "Quantum cascade lasers (QCLs) in biomedical spectroscopy," Chem. Soc. Rev. **46**(19), 5903–5924 (2017).
[34] N. S. Kapany and R. J. Simms, "Recent developments in infrared fiber optics[*]," Infrared Phys. **5**(2), 69–80 (1965).
[35] J. Wang, G. Wu, Z. Feng, J. Wang, Y. Wang, K. Jiao, X. Wang, S. Bai, P. Zhang, Z. Zhao, R. Wang, X. Wang, and Q. Nie, "Se-H-free $As_2Se_3$ fiber and its spectral applications in the mid-infrared," Opt. Express **30**(13), 24072 (2022).
[36] Y. Wang, K. Jiao, X. Liang, J. Jia, N. Li, S. Bai, X. Wang, Z. Zhao, Z. Liu, P. Zhang, S. Dai, Q. Nie, and R. Wang, "Fabrication of mid-IR As-Se chalcogenide glass and fiber with low scattering loss," J. Light. Technol. **42**(9), 3338–3345 (2024).
[37] V. Artyushenko, A. Bocharnikov, T. Sakharova, and I. Usenov, "Mid-infrared fiber optics for 1 — 18 μm range: IR-fibers and waveguides for laser power delivery and spectral sensing," Opt. Photon. **9**(4), 35–39 (2014).
[38] E. M. Dianov, V. G. Plotnichenko, G. G. Devyatykh, M. F. Churbanov, and I. V. Scripachev, "Middle-infrared chalcogenide glass fibers with losses lower than 100 db $km^{-1}$," Infrared Phys. **29**(2–4), 303–307 (1989).
[39] C. You, S. Dai, P. Zhang, Y. Xu, Y. Wang, D. Xu, and R. Wang, "Mid-infrared femtosecond laser-induced damages in $As_2S_3$ and $As_2Se_3$ chalcogenide glasses," Sci. Rep. **7**(1), 6497 (2017).
[40] S. Danto, M. Dubernet, B. Giroire, J.D. Musgraves, P. Wachtel, T. Hawkins, J. Ballato, and K. Richardson, "Correlation between native $As_2Se_3$ preform purity and glass optical fiber mechanical strength," Mater. Res. Bull. **49**, 250–258 (2014).
[41] W. H. Kim, V. Q. Nguyen, L. B. Shaw, L. E. Busse, C. Florea, D. J. Gibson, R. R. Gattass, S. S. Bayya, F. H. Kung, G. D. Chin, R. E. Miklos, I. D. Aggarwal, and J. S. Sanghera, "Recent progress in chalcogenide fiber technology at NRL," J. Non-Cryst. Solids **431**, 8–15 (2016).
[42] P. Lucas, A. A. Wilhelm, M. Videa, C. Boussard-Plédel, and B. Bureau, "Chemical stability of chalcogenide infrared glass fibers," Corros. Sci. **50**(7), 2047–2052 (2008).

[43] B. Hecht, B. Sick, U. P. Wild, V. Deckert, R. Zenobi, O. J. F. Martin, and D. W. Pohl, "Scanning near-field optical microscopy with aperture probes: Fundamentals and applications," J. Chem. Phys. **112**(18), 7761–7774 (2000).
[44] W. Bao, M. Melli, N. Caselli, F. Riboli, D. S. Wiersma, M. Staffaroni, H. Choo, D. F. Ogletree, S. Aloni, J. Bokor, S. Cabrini, F. Intonti, M. B. Salmeron, E. Yablonovitch, P. J. Schuck, and A. Weber-Bargioni, "Mapping local charge recombination heterogeneity by multidimensional nanospectroscopic imaging," Science **338**(6112), 1317–1321 (2012).
[45] S. Sabbatini, C. Conti, G. Orilisi, and E. Giorgini, "Infrared spectroscopy as a new tool for studying single living cells: Is there a niche?," Biomed. Spectrosc. Imaging **6**(3–4), 85–99 (2017).
[46] M. J. Baker, J. Trevisan, P. Bassan, R. Bhargava, H. J. Butler, K. M. Dorling, P. R. Fielden, S. W. Fogarty, N. J. Fullwood, K. A. Heys, C. Hughes, P. Lasch, P. L. Martin-Hirsch, B. Obinaju, G. D. Sockalingum, J. Sulé-Suso, R. J. Strong, M. J. Walsh, B. R. Wood, P. Gardner, and F. L. Martin, "Using Fourier transform IR spectroscopy to analyze biological materials," Nat. Protoc. **9**(8), 1771–1791 (2014).
[47] L. Vaccari, G. Birarda, L. Businaro, S. Pacor, and G. Grenci, "Infrared microspectroscopy of live cells in microfluidic devices (MD-IRMS): Toward a powerful label-free cell-based assay," Anal. Chem. **84**(11), 4768–4775 (2012).
[48] C. Gervillié-Mouravieff, C. Boussard-Plédel, J. Huang, C. Leau, L. A. Blanquer, M. B. Yahia, M. -L. Doublet, S. T. Boles, X. H. Zhang, J. L. Adam, and J. -M. Tarascon, "Unlocking cell chemistry evolution with operando fibre optic infrared spectroscopy in commercial Na(Li)-ion batteries," Nat. Energy **7**(12), 1157–1169 (2022).
[49] M. M. Amaral, C. G. Real, V. Y. Yukuhiro, G. Doubek, P. S. Fernandez, G. Singh, and H. Zanin, "In situ and operando infrared spectroscopy of battery systems: Progress and opportunities," J. Energy Chem. **81**, 472–491 (2023).
[50] J. -T. Li, Z. -Y. Zhou, I. Broadwell, and S. -G. Sun, "In-situ infrared spectroscopic studies of electrochemical energy conversion and storage," Acc. Chem. Res. **45**(4), 485–494 (2012).
[51] J. Kim, F. Zhao, S. Zhou, K. S. Panse, and Y. Zhang, "Spectroscopic investigation of the structure of a pyrrolidinium-based ionic liquid at electrified interfaces," J. Chem. Phys. **156**(11), 114701 (2022).
[52] J. Matsuura, A. Sheelam, and Y. Zhang, "Tandem supported, high metal-loading, non-PGM electrocatalysts for oxygen reduction reaction," APL Energy **2**(2), 026103 (2024).
[53] F. Zhao, S. Zhou, and Y. Zhang, "Ultrasensitive detection of hydrogen peroxide using $Bi_2Te_3$ electrochemical sensors," ACS Appl. Mater. Interfaces **13**(3), 4761–4767 (2021).
[54] E. Groppo, S. Rojas-Buzo, and S. Bordiga, "The role of *in situ*/*operando* IR spectroscopy in unraveling adsorbate-induced structural changes in heterogeneous catalysis," Chem. Rev. **123**(21), 12135–12169 (2023).
[55] C. Lamberti, A. Zecchina, E. Groppo, and S. Bordiga, "Probing the surfaces of heterogeneous catalysts by in situ IR spectroscopy," Chem. Soc. Rev. **39**(12), 4951 (2010).
[56] J. Zhao, X. Zhang, J. Xu, W. Tang, Z. Lin Wang, and F. Ru Fan, "Contact-electro-catalysis for direct synthesis of $H_2O_2$ under ambient conditions," Angew. Chem. Int. Ed. **62**(21), e202300604 (2023).
[57] Z. Wang, H. N. Fernández-Escamilla, J. Guerrero-Sánchez, N. Takeuchi, and F. Zaera, "Adsorption and reactivity of chiral modifiers in heterogeneous catalysis: 1-(1-naphthyl)ethylamine on Pt surfaces," ACS Catal. **12**(17), 10514–10521 (2022).

[58] L. K. S. Bonagiri, K. S. Panse, S. Zhou, H. Wu, N. R. Aluru, and Y. Zhang, "Real-space charge density profiling of electrode–electrolyte interfaces with angstrom depth resolution," ACS Nano **16**(11), 19594–19604 (2022).
[59] S. Zhou, K. S. Panse, M. H. Motevaselian, N. R. Aluru, and Y. Zhang, "Three-dimensional molecular mapping of ionic liquids at electrified interfaces," ACS Nano **14**(12), 17515–17523 (2020).
[60] K. S. Panse, H. Wu, S. Zhou, F. Zhao, N. R. Aluru, and Y. Zhang, "Innermost ion association configuration is a key structural descriptor of ionic liquids at electrified interfaces," J. Phys. Chem. Lett. **13**(40), 9464–9472 (2022).
[61] K. S. Panse, S. Zhou, and Y. Zhang, "3D mapping of the structural transitions in wrinkled 2D membranes: Implications for reconfigurable electronics, memristors, and bioelectronic interfaces," ACS Appl. Nano Mater. **2**(9), 5779–5786 (2019).
[62] L. K. S. Bonagiri, Z. Wang, S. Zhou, and Y. Zhang, "Precise surface profiling at the nanoscale enabled by deep learning," Nano Lett. **24**(8), 2589–2595 (2024).
[63] L. Lin, Y. Ge, H. Zhang, M. Wang, D. Xiao, and D. Ma, "Heterogeneous catalysis in water," JACS Au **1**(11), 1834–1848 (2021).
[64] M. Cortes-Clerget, J. Yu, J. R. A. Kincaid, P. Walde, F. Gallou, and B. H. Lipshutz, "Water as the reaction medium in organic chemistry: From our worst enemy to our best friend," Chem. Sci. **12**(12), 4237–4266 (2021).
[65] B. L. Dargaville and D. W. Hutmacher, "Water as the often neglected medium at the interface between materials and biology," Nat. Commun. **13**(1), 4222 (2022).
[66] P. M. Wiggins, "Role of water in some biological processes," Microbiol. Rev. **54**(4), 432–449 (1990).
[67] J. Li, J. Guo, and H. Dai, "Probing dissolved $CO_2$(aq) in aqueous solutions for $CO_2$ electroreduction and storage," Sci. Adv. **8**(19), eabo0399 (2022).
[68] N. Kitadai, T. Yokoyama, and S. Nakashima, "Temperature dependence of molecular structure of dissolved glycine as revealed by ATR-IR spectroscopy," J. Mol. Struct. **981**(1–3), 179–186 (2010).
[69] I. M. McIntosh, A. R. L. Nichols, K. Tani, and E. W. Llewellin, "Accounting for the species-dependence of the 3500 $cm^{-1}$ $H_2O_t$ infrared molar absorptivity coefficient: Implications for hydrated volcanic glasses," Am. Mineral. **102**(8), 1677–1689 (2017).
[70] H. Wijnja and C. P. Schulthess, "ATR–FTIR and DRIFT spectroscopy of carbonate species at the aged γ-Al2O3/water interface," Spectrochim. Acta A **55**(4), 861–872 (1999).
[71] A. E. Klingbeil, J. B. Jeffries, and R. K. Hanson, "Temperature-dependent mid-IR absorption spectra of gaseous hydrocarbons," J. Quant. Spectrosc. Radiat. Transfer **107**(3), 407–420 (2007).
[72] S. Yu. Venyaminov and F. G. Prendergast, "Water ($H_2O$ and $D_2O$) molar absorptivity in the 1000–4000 $cm^{-1}$ range and quantitative infrared spectroscopy of aqueous solutions," Anal. Biochem. **248**(2), 234–245 (1997).
[73] J. Kim, U. W. Schmitt, J. A. Gruetzmacher, G. A. Voth, and N. E. Scherer, "The vibrational spectrum of the hydrated proton: Comparison of experiment, simulation, and normal mode analysis," J. Chem. Phys. **116**(2), 737–746 (2002).
[74] A. B. McCoy, "The role of electrical anharmonicity in the association band in the water spectrum," J. Phys. Chem. B **118**(28), 8286–8294 (2014).
[75] X. Lu, C. Zhu, Z. Wu, J. Xuan, J. S. Francisco, and H. Wang, "In situ observation of the pH gradient near the gas diffusion electrode of $CO_2$ reduction in alkaline electrolyte," J. Am. Chem. Soc. **142**(36), 15438–15444 (2020).

[76] H. H. Heenen, H. S. Pillai, K. Reuter, and V. J. Bukas, "Exploring mesoscopic mass transport effects on electrocatalytic selectivity," Nat. Catal. **7**(7), 847–854 (2024).
[77] B. Chen, Y. Xia, R. He, H. Sang, W. Zhang, J. Li, L. Chen, P. Wang, S. Guo, Y. Yin, L. Hu, M. Song, Y. Liang, Y. Wang, G. Jiang, and R. N. Zare, "Water–solid contact electrification causes hydrogen peroxide production from hydroxyl radical recombination in sprayed microdroplets," Proc. Natl. Acad. Sci. **119**(32), e2209056119 (2022).
[78] Y. Zhang, S. Zhan, K. Liu, M. Qiao, N. Liu, R. Qin, L. Xiao, P. You, W. Jing, and N. Zheng, "Heterogeneous hydrogenation with hydrogen spillover enabled by nitrogen vacancies on boron nitride-supported Pd nanoparticles," Angew. Chem. Int. Ed. **62**(9), e202217191 (2023).
[79] L. Khalafi, N. Nikzad, A. Alhajeri, B. Bacon, K. Alvarado, and M. Rafiee, "Electrochemistry under microscope: Observing the diffusion layer and measuring diffusion coefficient," J. Chem. Educ. **100**(10), 4056–4061 (2023).
[80] R. C. Engstrom, M. Weber, D. J. Wunder, R. Burgess, and S. Winquist, "Measurements within the diffusion layer using a microelectrode probe," Anal. Chem. **58**(4), 844–848 (1986).
[81] S. M. Oja and B. Zhang, "Imaging transient formation of diffusion layers with fluorescence-enabled electrochemical microscopy," Anal. Chem. **86**(24), 12299–12307 (2014).
[82] X. Lang, L. Shi, Z. Zhao, and W. Min, "Probing the structure of water in individual living cells," Nat. Commun. **15**(1), 5271 (2024).
[83] J. T. King, E. J. Arthur, C. L. Brooks, and K. J. Kubarych, "Crowding induced collective hydration of biological macromolecules over extended distances," J. Am. Chem. Soc. **136**(1), 188–194 (2014).